\documentclass[11pt, a4paper]{article}
\pdfoutput=1 
\usepackage{jcappub}

\usepackage[T1]{fontenc} 


\usepackage[utf8]{inputenc}
\usepackage{subcaption}
\usepackage{setspace}
\usepackage{color}
\usepackage{epstopdf}
\usepackage{makeidx} 
\usepackage{lmodern}
\usepackage{packages} 

\usepackage[center]{titlesec}
\usepackage{bm}
\usepackage{amsmath}                    
\usepackage{graphicx}
\usepackage{amssymb}
\usepackage{journals}
\usepackage[export]{adjustbox}

\usepackage[T1]{fontenc}
\usepackage[utf8]{inputenc}
\usepackage{color}
\usepackage{booktabs}
\usepackage{units}
\usepackage{esint}
\usepackage{braket}

\usepackage{caption}
\usepackage{subcaption}
\usepackage[export]{adjustbox}
\pdfpageheight\paperheight
\pdfpagewidth\paperwidth

\makeatletter

\usepackage{graphics}

\usepackage{setspace}
\usepackage{color}
\usepackage{epstopdf}
\usepackage{setspace}
\usepackage{makeidx}

\usepackage{anysize}
\marginsize{2.5cm}{2cm}{1cm}{3.5cm}

\def \mpc {\mbox{\rm Mpc}}

\def\be{\begin{equation}}
\def\ee{\end{equation}}

\title{Future constraints on the gravitational slip with the mass profiles of galaxy clusters}

\author[1]{Lorenzo Pizzuti,}
\author[2]{Ippocratis D. Saltas,}
\author[3]{Santiago Casas,}
\author[4]{Luca Amendola,}  
\author[5]{Andrea Biviano}

\affiliation[1]{Dipartimento di Fisica, Sezione di Astronomia, Universit\`a di Trieste,\\ Via Tiepolo 11, I-34143 Trieste, Italy}
\affiliation[2]{CEICO, Institute of Physics of the Czech Academy of Sciences, Na Slovance 2, 182 21 Praha 8, Czechia}
\affiliation[3]{Institut f\"ur Theoretische Physik, Universit\"at Heidelberg,\\ Philosophenweg 16, D-69120 Heidelberg, Germany}
\affiliation[4]{AIM, CEA, CNRS, Universit\'e Paris-Saclay, Universit\'e Paris Diderot,\\ Sorbonne Paris Cit\'e, F-91191 Gif-sur-Yvette, France}
\affiliation[5]{INAF-Osservatorio Astronomico di Trieste, via G. B. Tiepolo 11, 34143, Trieste, Italy}

\abstract{The gravitational slip parameter is an important discriminator between large classes of gravity theories at cosmological and astrophysical scales. In this work we use a combination of simulated information of galaxy cluster mass profiles, inferred by Strong+Weak lensing analyses and by the study of the dynamics of the cluster member galaxies, to reconstruct the gravitational slip parameter $\eta$ and predict the accuracy with which it can be constrained with current and future galaxy cluster surveys. Performing a full-likelihood statistical analysis, we show that galaxy cluster observations can constrain $\eta$ down to the percent level already with a few tens of clusters. We discuss the significance of possible systematics, and show that the cluster masses and numbers of galaxy members used to reconstruct the dynamics mass profile have a mild effect on the predicted constraints.}
\begin{document}
\maketitle

\section{Introduction}
The dark energy problem has challenged our understanding of gravity at large scales in the Universe. To explain it, a variety of different theoretical scenarios have been proposed based on the introduction of new degrees of freedom beyond the standard paradigm of General Relativity (GR) (for a review see e.g \cite{Clifton:2011jh, Amendola2010}). The recent impressive measurement of the speed of gravitational waves provided the most stringent test on viable modified gravity theories so far, with direct implications for cosmology and the large scale structure of the Universe. However, understanding the nature of the late--time acceleration of the Universe calls for further, complementary observational information, that could shed light on the viability of the surviving theory space. 

From a phenomenological point of view, modifications of GR leave two key imprints at the level of large scale structures in the Universe. In particular, they affect the effective strength of gravity, impacting the clustering of matter at large scales, while they may also impact the way gravity interacts with light, leaving a characteristic imprint on weak lensing observables. Both effects, however, could be equally due to new gravitational degree(s) of freedom that do not modify gravity, such as minimally-coupled scalar-tensor models (quintessence, k--essence), so that they cannot be used to detect genuine departures from GR. A crucial  parameter in this regard is the gravitational slip $\eta$ defined as the ratio between the effective gravitational coupling of light to that of matter (to be defined later on). In GR and in models where the new degree of freedom is minimally-coupled to gravity one has $\eta = 1$, while an $\eta \neq 1$ would be the smoking gun for a  departure from GR (see e.g \cite{Amendola:2012ys}) at late times, at least as long as dark matter can be described as a perfect fluid \footnote{We notice here that, free streaming neutrinos can also be a source of gravitational slip. However, at late times in the cosmological evolution their contribution can be neglected.}. Crucially, $\eta$ can be reconstructed in a model-independent manner by combining galaxy kinematics and weak lensing observations \cite{Amendola:2012ky,Motta:2013cwa}, while it is directly linked to the properties of gravitational wave propagation at any scale \cite{Saltas:2014dha,Sawicki:2016klv,Amendola:2017orw}. Therefore, its reconstruction from observations is key in testing departures from GR and/or distinguishing amongst large classes of gravity models at both cosmological and astrophysical scales. 

Galaxy clusters are ideal laboratories in this regard, as they can provide us with simultaneous information about the local gravitational potentials sourcing the dynamics of their member galaxies as well as the lensing of light, which can be combined to reconstruct $\eta$. In this sense, tests of gravity based on $\eta$ can be understood as a consistency check between the cluster's mass profile as inferred from kinematical and lensing observations respectively. Current and future galaxy cluster surveys observing at different, complementary bands, such as Euclid, Dark Energy Survey (DES), XMM-Newton, or eROSITA \footnote{http://www.darkenergysurvey.org/, http://www.mpe.mpg.de/heg/www/Projects/erosita/index.php, http://sci.esa.int/xmm-newton/43133-slew-survey-and-catalogue/. For a review on future surveys see \cite{Futurecluster}.}  aim to provide observations of hundreds of clusters with accurate mass determination. This work adds to previous efforts to constrain gravity with galaxy clusters in 
various theoretical setups, e.g in \cite{Sakstein:2016ggl, Wilcox:2016guw, Wilcox:2015kna, Terukina:2013eqa}, and it is complementary to conceptually similar reconstructions of $\eta$ using redshift space distortions, galaxy-galaxy lensing and the background expansion \cite{Pinho:2018unz}. 

The {\it goal} of the present work is two-fold: On the one hand, we aim at predicting the potential statistical accuracy of future constraints on $\eta$ from galaxy cluster kinematics and lensing, while at the same time understanding their dependence on key factors such as the number of member galaxies and the mass of the clusters in the sample. Our analysis is based on the dynamical mass profile reconstructions, performed with the \emph{MAMPOSSt} method of Ref. \cite{Mamon01}, of a set of simulated spherically symmetric isolated clusters. This approach will give us enough flexibility to investigate how different cluster masses and densities of tracers affect our constraints. For the complementary lensing information we will rely on the lensing mass profile inferred by Refs. \cite{UmetsuMACS,Umetsu16,Caminha16} on the massive, relaxed galaxy cluster MACS 1206, which belongs to a sample of 20 X-ray selected objects analysed within the CLASH/ CLASH-VLT collaborations (refs. \cite{Postman01,Rosati1}).

It should be noted that theories modifying gravity typically introduce a characteristic time and scale dependence on the gravitational slip parameter $\eta = \eta(t,k)$, which is often captured through appropriate parametrisations. Unlike previous constraints of gravity \cite{Sakstein:2016ggl, Wilcox:2016guw, Wilcox:2015kna, Terukina:2013eqa} based on particular gravity setups, in our analysis we will make no assumption on the particular functional dependence of $\eta$ on time and scale, the only assumption being that of a Navarro-Frenk-White parametrisation for the cluster mass profile. In this regard, our constraints will be sufficiently general -- any departures from the expectation $\eta=1$ is incorporated in a difference between the mass profile parameters inferred by dynamics and lensing probes, which can then be translated into constraints on the parameter space of particular theories of gravity.

We show that already with $\sim 15$ clusters we can get at $1\sigma$ an accuracy of 
$~ \sim 5 \%$ for a scale-independent and $~ \sim 20\%$ for a scale-dependent $\eta$.
The accuracy is significantly improved down to the few percent level for both cases when extrapolated to $\sim 75$ clusters. We investigate the effect of varying the number of tracers (i.e. member galaxies) in the dynamics analysis, the dependence of constraints on the cluster mass, as well as the uncertainties in the simulated lensing probability distribution. 

We structure the paper as follows: Section \ref{sec:defs} lays down the main assumptions and definitions of our analysis, while in Section \ref{sec:Theory} we introduce the essential theoretical framework. A detailed exposition of all our simulations is presented in Section \ref{sec:Dynamics}, and finally, the results of the statistical analysis are explained in Section \ref{sec:results}. The main results are summarised in the last section of the paper, while a summary of the predicted constraints can be found in Table \ref{tab:res}.

\section{Essential assumptions and definitions} \label{sec:defs}
An essential step towards forecasting the potential of future galaxy-cluster surveys to constrain the gravitational slip parameter is having a handle upon a measurement of  clusters dynamical and lensing masses, as we explain in more detail in Section \ref{sec:Theory}. In this regard, we will combine information from a set of reliable simulated data based on current and future observations. The analysis will inevitably rely on certain fundamental simplifications and assumptions about the structure and dynamics of the clusters and the underlying cosmology, which is important to lay out before we proceed. They can be summarized as follows:
\\

i. The clusters will be modelled as spherical inhomogeneities in a complete equilibrium, that is, in thermal and hydrostatic equilibrium. In other words, the clusters will be assumed to be relaxed, with their mass distribution and gravitational potentials being time-independent.
Although from an observational viewpoint it might be challenging to identify truly relaxed clusters, this simplification will allow us to avoid the introduction of systematics and/or degeneracies that we would otherwise encounter when dealing with a non-relaxed cluster. We comment on the issue of departures from relaxation in Section \ref{sec:Theory}. Additional assumptions related to the anisotropy of the cluster's velocity field are also explained in Section \ref{sec:jeans}. 
\\ 

ii. The density distribution of the cluster will be dominated by dark matter, which we will model by a Navarro-Frenk-White (NFW) profile. The member galaxies will be assumed to be collissionless and non-interacting point particles moving along geodesics and populated according to the same profile. The validity of this assumption is ensured within the linear regime, whereas it is expected to break down as soon as non-linear galactic interactions become important. These are however rare, given the high speed of
galaxy-galaxy encounters in clusters.
\\ 

iii. The background cosmology will be that of the $\Lambda$CDM model, while large-scale structures will be sourced by linear scalar fluctuations according to $ds^2 = -(1+ \Psi(t,{\bf x}))dt^2 + a(t)^2 (1 +\Phi(t, {\bf x} ))d {\bf x}^2$. Under this convention, we define the gravitational slip parameter as $\eta \equiv \Phi/\Psi$.
\\

iv. The functional form of the gravitational slip parameter $\eta = \eta(t,k)$ as a function of time and scale depends on the particular gravity action under consideration. Here, we shall make no a priori assumption about the actual gravity model, proceeding in a sufficiently general manner considering two distinct cases: A case where $\eta$ is assumed to be scale-independent throughout our clusters sample, and another one where it depends on the cluster's spatial scale through the characteristic scale $r_s$ of the NFW profile function. These two cases encompass the predictions of scalar--tensor theories at different (subhorizon) scales, as explained in Section \ref{sec:Theory}.

\section{Gravitational slip as a consistency condition on the cluster's dynamical and lensing mass}\label{sec:Theory}
Testing the gravitational slip at the galaxy cluster level can be understood as a consistency check between the lensing and dynamical mass of the cluster, $M_{dyn}$ and $M_{lens}$ respectively. In GR, $M_{dyn}= M_{lens}$, which is no longer true as soon as gravity is modified \footnote{Our definition of modified gravity is that of \cite{Amendola:2017orw}, defined as any theory modifying the dynamics of the genuine degrees of freedom of GR, that is the two polarisations of the spin-two field.}.To understand why this is so, let us assume a cluster to be in a state of dynamical equilibrium with the total gravitational potential $\Psi$. If interactions and dissipative effects are negligible,  the cluster member galaxies acts like collisionless tracers of the gravitational potential and their dynamics is governed by the Jeans equation: 
\begin{equation}\label{eq:jeans}
\frac{\partial (\nu\sigma_r^2)}{\partial t}+2\beta(r)\frac{\nu\sigma^2_r}{r}=-\nu(r)\frac{\partial \Psi}{\partial r} \, ,
\end{equation}
with $\nu(r)$ the number density of tracers, $\sigma^2_r$ the velocity dispersion along the radial direction and $\beta \equiv 1-(\sigma_{\theta}^2+\sigma^2_{\phi})/2\sigma^2_r$ the velocity anisotropy profile.
The gravitational potential $\Psi$ is connected to the matter density distribution through the Poisson equation:
\begin{equation}
\nabla^2 \Psi = 4 \pi G \rho_m \, ,
\end{equation}
where $\rho_m$ stands for the total mass density of the cluster at a given radius $r$. The total density is assumed to be dominated by the dark matter component, i.e $\rho_m = \rho_{\text{DM}} + \rho_{\text{gas}}+\rho_{\text{gal}} \simeq  \rho_{\text{DM}} $, as baryons provide a negligible contribution. The total mass of the (spherical) cluster with total radius $R$ is given by the mass conservation equation:
\begin{equation}
M_{\text{tot}}(R)=4 \pi \int^{R}r^{2}\rho_m(r)\mbox{d}r \, . \label{eq:Mass}
\end{equation}
In order to model the dark matter density we will follow a parametric approach through the use of a NFW profile \cite{navarro97}, which  describes the distribution of collisionless Cold Dark Matter (CDM) in spherical cluster-size halos in an equilibrium configuration, defined through
\begin{align}
\rho_{\text{NFW}}(r) = \frac{\rho_\text{s}}{(r/r_s)(1+r/r_s)^2} \, ,
\end{align}
with $\rho_s$ and $r_s$ two free parameters to be fit from observations. The NFW profile is in very good agreement with N-body simulations. In particular, $r_{s}$ is a characteristic scale radius, corresponding to the distance from the center of the cluster where $d(\ln \rho)/d \ln r = -2$, while $\rho_s$ is  defined as
\be \label{eq:NFWdens}
\rho_s=\rho_c(z)\frac{\Delta_{vir}}{3}\frac{c^3}{\log(1+c)-c/(1+c)} \, .
\ee
Here, $\rho_c(z)=3H^2(z)/8\pi G$ is the critical density of the Universe at redshift $z$, and $c \equiv r_{vir}/r_s$ is the so--called \emph{concentration} parameter, while $r_{vir}$ denotes the radius enclosing a mean overdensity $\Delta_{vir}$ times the critical density of the Universe. In this paper we will use $\Delta_{vir}=200$ (and so, $r_{vir}=r_{200}$), which is close to the density at virialization predicted by the spherical collapse model in a EdS universe $\Delta_{sc}=178$. \\
By inserting equation \eqref{eq:NFWdens} into \eqref{eq:Mass} we obtain the halo mass as a function of radius as
\be \label{eq:NFW}
M(r) = M_{200} \frac{\ln(1 + r/r_{s}) - (r/r_{s})(1+r/r_{s})^{-1}}{\ln(1+c) - c/(1+c) } \, ,
\ee
with $M(r_{200})= M_{200}$ the total mass enclosed in a sphere of radius $r_{200}$. Despite its relative simplicity,  the NFW profile has been widely adopted in the literature to describe the mass profile of gravitationally bound structures, as it has been shown to provide a generally good fit to observational data (e.g. refs. \cite{Biviano01,Umetsu16}). The choice of a parametric description of the density perturbations has the noticeable advantage to simplify the extension of the analysis to specific modified gravity theories, where departures from GR can be easily incorporated in additional parameters in the model, as we will discuss in more detail later on. 
Now, assuming spherical symmetry the Poisson equation can be integrated to yield
\begin{equation}
\Psi(r)=G\int^r_{r_0} \frac{ds}{s^2}M_{\text{dyn}}(s) \, . \label{eq:dynamical-mass}
\end{equation}
The above equation provides a definition for a cluster dynamical mass, that is, the mass one would infer through galaxy kinematics measurements, related to the Newtonian potential $\Psi$. 
We can derive a similar definition for the lensing mass, considering that light geodesics respond to the weak lensing potential $\Phi+\Psi$, described by the Poisson-type equation through
\begin{equation}
\nabla^{2}(\Phi+\Psi)=8\pi G\rho_{\text{lens}} \, , \label{eq:wl-pot-poisson}
\end{equation}
which under the assumption of spherical symmetry becomes
\begin{equation}
M_{\text{lens}}=\frac{r^{2}}{2G} \frac{d}{dr}(\Phi + \Psi) \, .
\end{equation}
Equations \eqref{eq:wl-pot-poisson} and (\ref{eq:dynamical-mass}) allow us to solve for the potential $\Phi$,
\begin{equation}
\Phi(r)=G\int^r\frac{ds}{s^2}\left[2M_{\text{lens}}(s)-M_{\text{dyn}}(s)\right].  
\end{equation}
Combining the expressions for the potentials $\Phi$ and $\Psi$ we can compute the ratio $\Phi / \Psi$, to arrive to a relation for the gravitational slip parameter $\eta$ in terms of the lensing and dynamical mass of the cluster as,
\begin{equation}
\eta(r) =\frac{\int^r \frac{ds}{s^2}\left[2M_{\text{lens}}(s)-M_{\text{dyn}}(s)\right]}{\int^r \frac{ds}{s^2}M_{\text{dyn}}(s)} \, . \label{eq:eta}
\end{equation}
Equation (\ref{eq:eta}) will be our starting point in predicting the constraints on $\eta$ at the cluster level. Notice that, in GR it is $M_{\text{dyn}} = M_{\text{lens}}$ and $\eta = 1$. 

In the usual description of $\eta$ in general classes of gravity models, one usually works in momentum (Fourier) space, where the scale dependence is controlled through the Fourier wavenumber $k$. For example, for the popular class of the general Horndeski scalar-tensor theories in the sub-horizon quasi-static regime $\eta$ acquires the form
\begin{align}
\eta = h_2 \cdot \frac{1 + h_{4} k^2}{1 + h_{5} k^2} \, ,
\end{align}
with the functions $h_{i} = h_{i}(t)$ defined in \cite{DeFelice:2011hq}. For the particular case of Brans--Dicke/$f(R)$ gravity they depend on the scalaron Compton wavelength. However, since our simulations, and hence our analysis, will be in real space it is essential to discuss how our parametrisation relates to the Fourier space analyses. To get the real space $\eta$ one would have to express the linearised equations for $\Phi$ and $\Phi + \Psi$ in real space, assuming a parametrisation for the density profile. In general, we expect the resulting expression for for generic theories of gravity to acquire the following schematic form
\be
\eta(r,t) = f_{1}(t) \left(1 + f_{2}(r,t) \right) \, , \label{eq:eta-realspace}
\ee
with GR corresponding to the limit $f_{1} \to 1, f_{2} \to 0$. A scale--independent $\eta$ is sourced only by $f_{1}$, which is in principle time-dependent -- in our analysis, we will confine ourselves to a fixed redshift, which enforces $\eta$ to be a constant for this case. Without the assumption of a model-dependent parametrisation our approach will be rather phenomenological, yet sufficiently general to encompass any gravity theory. In particular, in view of (\ref{eq:eta}) and the assumption of a NFW profile for the matter distribution, it is easy to see that $\eta$ will be sourced through a difference of the inferred parameters $r_{200}, r_{s}$ from lensing and dynamics respectively, i.e
\be
\eta(r; r_{200}^{D}, r_{s}^{D}) \neq \eta(r; r_{200}^{L}, r_{s}^{L}) \, ,
\ee
unless $(r_{200}^{D}, r_{s}^{D}) =( r_{200}^{L}, r_{s}^{L})$. From an observational viewpoint, the existence of some new degree of freedom sourcing $\eta \neq 1$, will manifest itself as a tension between the inferred NFW parameters from lensing and dynamics respectively, when trying to reconstruct the mass distribution of the cluster. In this regard, the strategy in our analysis will be as follows: the synthetic clusters will be constructed with different initial conditions implying different distribution for the inferred NFW profiles from dynamics, $(r_{200}^{D}, r_{s}^{D})$. The resulting distributions from the dynamics will be in principle different from those from lensing, hence sourcing gravitational slip under equation (\ref{eq:eta}). The derived constraints on $\eta$ can then be mapped to specific theories of gravity. Below, we explain the simulations and statistical analysis in detail, before we dwell upon the predicted constraints. 

\section{Galaxy cluster simulations and statistical analysis} \label{sec:Dynamics}
Predicting the constraints on $\eta$ from current or future galaxy cluster observations requires a sample of galaxy clusters along with predictions for the dynamical and lensing mass of each member of the sample according to equation \eqref{eq:eta}. The {\it principal aim of our analysis} will be to estimate the number of reliable lensing and dynamics mass reconstructions, which will be available from upcoming imaging and spectroscopic surveys, needed  in order to constrain $\eta$ at the {\it percent level} of statistical accuracy. For this purpose, we will neglect the impact of systematic effects in mass profile determinations, i.e. deviation from dynamical relaxation and spherical symmetry, as well as errors introduced by non-suitable parametrisations of the density profile. Therefore, we will account only for the statistical power of the presented method.  An extensive analysis of the impact of systematics in constraining modified gravity models with cluster mass profiles will be performed in a forthcoming work (Pizzuti et al., in prep.).

The analysis carried out in \cite{Pizzuti16} on the X-ray selected CLASH cluster MACS 1206 has shown that accurate dynamics and lensing mass profile reconstructions allow to constrain $\eta(r)$ down to $30\%$ statistical uncertainties with a single galaxy cluster when the systematic effects are under control. 

In reconstructing a cluster mass profile from the dynamics of the member galaxies we proceed in {\it two steps} as follows: we first generate a synthetic sample of 15 equilibrium configurations of isolated, self-gravitating systems populated by collisionless point-like tracers according to a spherical NFW profile. Each synthetic cluster is constructed with a different characteristic mass and scale, corresponding to different choices for the $r_{200}$ and $r_s$ parameters in the NFW profile, covering two orders of magnitude in cluster masses (see Table \ref{eq:tracers}). 
An interesting outcome of our analysis, as we will see in Section \ref{sec:results}, is that the constraints do not show an appreciable dependence on halo masses and concentrations. The velocity dispersion of the tracers is assigned following the solution of the Jeans equation, as explained in Section \ref{sec:jeans}. The particular number of 15 synthetic halos is chosen as a fair compromise between computing time and a statistically representative sample of cluster masses and concentrations range. The second step, discussed in Sec. \ref{sec:MAM}, consists in inputing positions and velocities of the halo's particles into the \emph{MAMPOSSt} procedure of Ref. \cite{Mamon01}, to perform a maximum likelihood fit of the cluster mass profile based on the tracers velocity field. 
For the lensing part we simulate the probability distribution $P_L(r_s,r_{200})$ relying on the results of the analysis of Ref. \cite{UmetsuMACS}, which derived the strong+weak lensing mass profile for the galaxy cluster MACS 1206. The profile is parametrized as a NFW model  with $\sim 6\%$ and $30\%$ uncertainties on the mass profile parameters $r_{200}$ and $r_s$ respectively. Moreover, it was found that the joint probability distribution of the NFW parameters is almost Gaussian with a non-zero covariance between $r_s$ and $r_{200}$. 

We thus use a bivariate Gaussian distribution,
\begin{displaymath}
 P_L(r_s,r_{200})=\frac{1}{2\pi\sigma_{r_s}\sigma_{r_{200}}\sqrt{1-\rho^2}}\exp\left\{-\frac{1}{2(1-\rho^2)}\left[\frac{(r_s-\bar{r}_s)^2}{\sigma^2_{r_s}}+\right.\right.
\end{displaymath}
\begin{equation}
\left.\left.+\frac{(r_{200}-\bar{r}_{200})^2}{\sigma^2_{r_{200}}}-\frac{2\rho(r_s-\bar{r}_s)(r_{200}-\bar{r}_{200})}{ \sigma_{r_s}\sigma_{r_{200}}} \right]\right\} \, ,
\end{equation} 
assumed to be centered on the best fit values of the NFW parameters $\bar{r}_s,\bar{r}_{200}$ given by the dynamics reconstruction, as we will explain in Section \ref{sec:results}. Here $\rho$ indicates the correlation. In general, the virial radius $r_{200}$, which is related to the total cluster mass, can be constrained much better than the characteristic radius of the halo mass profile, expressed in terms of $r_s$. As shown e.g. in Table 2 of Ref. \cite{Umetsu16}, typical uncertainties on the scale radius given by a lensing probe are of the order of $\sim30\div40\%$, while $r_{200}$ can be recovered up to $\sim 5\div10\%$. For our reference analysis we consider $\sigma_{r_{200}}=0.07\times r_{200}$ and $\sigma_{r_s}=0.30\times r_s$, while for the correlation in  $P_L(r_s,r_{200})$ we adopt the value $\rho=0.67$, found by fitting a bivariate Gaussian on the posterior distribution of Ref. \cite{UmetsuMACS}. We further discuss the robustness of our results when varying both correlation and variances in the lensing distribution.

\subsection{Synthetic galaxy clusters} \label{sec:jeans}
Here we discuss the procedure for the generation of the simulated clusters and we list the main properties of the 15 produced synthetic halos we will use in our analysis. Under the assumption of dynamical relaxation, member galaxies in clusters act as collisionless tracers of the total gravitational potential and their dynamics is governed by the Jeans equation. In order to avoid systematic effects induced by lack of equilibrium in the mass profile determination, we generate each halo as an isolated system of collisionless particles assuming spherical symmetry, dynamical relaxation and 3D-Gaussian distribution of the velocities of the tracers (galaxies).
Given a $\Lambda$CDM background with $H_0=70\,\textrm{km}\,\textrm{Mpc}^{-1}\textrm{s}^{-1}$, $\Omega_m=0.3$ and $\Omega_{\Lambda}=0.7$\footnote{Note that the background cosmology enters only in the definition of $M_{200}$ for each cluster.}, we populate each halo at redshift $z=0$  according to a NFW density profile, eq. \eqref{eq:NFW}, up to $\sim9$ viral radii. 

It is natural to expect that the accuracy of the cluster's reconstructed dynamical mass will depend on the total number of its galaxy members within $r_{200}$. For this purpose, we perform our analysis for two different choices in this regard. In the first case, all 15 synthetic clusters are generated with the same number of tracers equal to 1000, an optimistic, although not unrealistic choice, given the expectation of the number of cluster galaxies with accurate measured spectroscopic redshift from present and future surveys. In the second case we assigned the number of tracers assuming that it is dependent on the cluster mass, requiring a minimum of 300 particles within $r_{200}$ for the less massive halo in the sample. In particular, for the total number of tracers within $r_{200}$ of the $i$-th cluster $N^{(i)}(r_{200})$ we use
\be
N_g^{(i)}(r_{200})=
\left\{
\begin{array}{lll}
300    &  & \; \text{$i=1$} \, , \\
\left( M^{(i)}_{200}/M^{(1)}_{200} \right) \cdot \frac{N_c}{i}  &   &  \; \text{$i>1$} \,. \label{eq:tracers}
\end{array}
\right.
\ee
We choose $N_c = 375$ in order to guarantee a maximum number of tracers less than 1400 for the most massive cluster. It is worth to notice that the choice of eq. \eqref{eq:tracers} is totally arbitrary and still an optimistic simplification, since the number of spectroscopically confirmed member galaxies in clusters strongly depends on several factors, such as observational time, completeness of the sample (see e.g. Ref. \cite{Biviano01}) and contamination from field galaxies.  The resulting $N_{g}(r_{200})$ are displayed in the fifth column of Table \ref{tab:perfc}.

The end result of the sample consists of 15 objects with masses spawning a range from $[5.1\cdot10^{13}\,M_{\odot} \div2.9\cdot10^{15}\,M_{\odot}]$. For each halo, we determine the scale radius $r_s$ according to the $c$-$M$ relation (extrapolated at $z=0$) of Ref.\cite{Merten14} \footnote{See also Ref. \cite{Duffy08}.}, which analyzed 19 of the X-ray selected clusters of the CLASH sample at $z \sim 0.19 - 0.89$ to find
\begin{equation}\label{eq:clash_cmrel}
c(M_{200},z)=A\times\left(\frac{1.37}{1+z}\right)^B \cdot \left(\frac{M_{200}}{8\times10^{14}M_{\odot}/h}\right)^K \,\, ,
\end{equation}
with
\begin{equation}\label{eq:parclash}
 A = 3.66 \pm 0.16,\,\,\,\,\, B = - 0.14\pm0.52,\,\,\,\,\,\, K= -0.32 \pm 0.18 \,\,.
\end{equation}
We have randomly sampled the value of the parameters $A,B$ and $K$ for every halo assuming that they are Gaussian-distributed around the central value and standard deviation of \eqref{eq:parclash}. The characteristic end-values for each synthetic cluster are summarized in Table \ref{tab:perfc}.
\begin{table}[h]
\centering
\begin{tabular}{|c|c|c|c|c|}
\hline
 {\bf Cluster ID} &  ${\bf r_{200}}\,[\text{Mpc}]$ & ${\bf r_s} \,[\text{Mpc}]$ & ${\bf r_{\nu} }\,[\text{Mpc}]$  &{\bf \# of tracers} ${\bf N_g(r_{200})}$ \\\hline
\hline
1&	0.750&	0.185&	0.169&	300	\\	
2&	0.900&	0.094&	0.097&	324\\
3&	1.050&	0.177&	0.178&	343	\\
4&	1.200&	0.133&	0.144&	384	\\
5&	1.350&	0.231&	0.219	&	437\\
6&	1.500&	0.258&	0.256	&	500\\
7&	1.650&	0.227&	0.252	&	570\\
8&	1.800&	0.360&	0.387	&	648\\
9&	1.950&	0.511&	0.466	&	732\\
10&	2.100&	0.515&	0.473	&	823\\
11&	2.250&	0.539&	0.503	&	920\\
12&	2.400&	0.614&	0.641	&	1024\\
13&	2.550&	0.848&	0.807	&	1133\\
14&	2.700&	0.993&	1.005	&	1249\\
15&	2.850&	1.112&	1.117	&	1371\\
\hline
\end{tabular}
\caption[Parameters of the synthetic halos used in the analysis]{\label{tab:perfc} A summary with the characteristics of the synthetic clusters sample used in our analysis ordered according to increasing mass. The second column shows the parameters $r_{200}$ for each cluster, sampled with a fixed step size of $0.15\,\mpc$, while the third column the corresponding $r_s$ parameter derived according to the empirical concentration-mass relation \eqref{eq:clash_cmrel}. The fourth column shows the scale radius $r_{\nu}$, obtained by fitting the projected number density profile from the phase space with a projected NFW model, which is additionally required in the \emph{MAMPOSSt} procedure. The fifth column denotes the number of tracers within a sphere of radius $ r = r_{200}$, computed according to eq. \eqref{eq:tracers}.}
\end{table}

\begin{figure}
\hspace{-1.5cm}\includegraphics[width=1.1\linewidth]{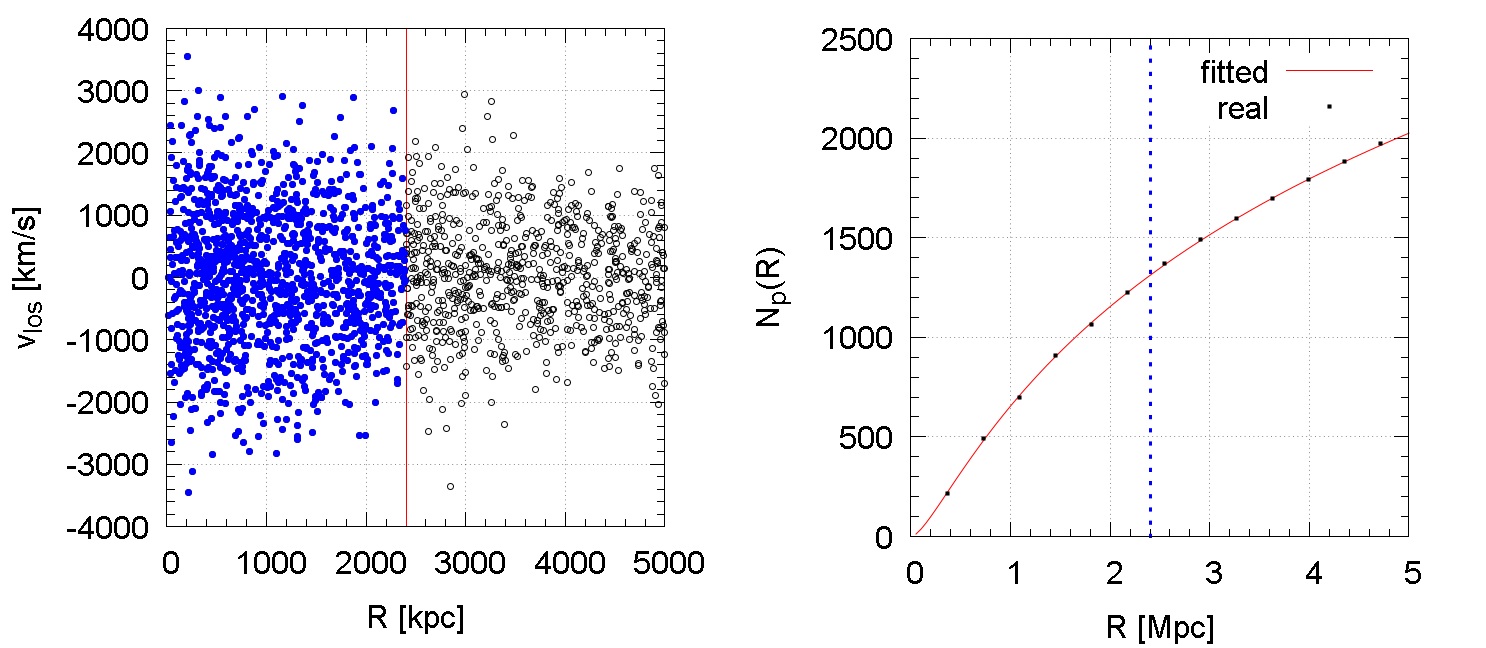}
        \caption{\label{fig:pps}{\it Left:} Projected phase space $(R, v_z=v_{los})$ for one of the synthetic clusters (ID 12) generated with $N(r_{200})=1024$. Each point indicates a cluster member; the red vertical line denotes the virial radius $r_{200}$. With blue points we mark the tracers used in the dynamical mass profile reconstruction through the \emph{MAMPOSSt} procedure. {\it Right:} Cumulative projected number density profile of the tracers as a function of projected radius $R$ (black points). The red solid curve shows the best fit projected NFW profile with scale radius $r_{\nu}$ indicated in Table \ref{tab:perfc}; vertical blue dotted line corresponds to $R=r_{200}$. }
\end{figure}
In order to better explore the dependency of the derived constraints on the density of tracers used in the dynamics analysis, we further consider an additional synthetic cluster with  $r_{200}=1.96\,\mpc$, $r_s=0.27\, \mpc$  generated six times by varying the number of particles $N_g(r_{200})$ from $50$ to $2000$.

To each particle at a radial distance $r$ from the center we assign a velocity whose components in spherical coordinates are Gaussian-distributed with a squared dispersion $\vec{\sigma}^2_{\mathbf{r}} (r) = \{ \sigma^2_r(r), \sigma^2_{\theta}(r), \sigma^2_{\phi}(r)\}$, given by the solution of the (spherical) Jeans equation, eq. \eqref{eq:jeans}. Note that the velocity anisotropy profile $\beta(r)$ is in principle unknown, and one of the major sources of systematics in the reconstruction of cluster mass profile with the analysis of the dynamics of member galaxies.  Different methods have been developed in the literature to reconstruct $\beta(r)$, e.g through fitting suitable parametrised profiles, or non-parametric techniques based on inversions of the Jeans equation (e.g. refs. \cite{Binney1982,Host2}), however, the latter generally requires additional information and assumptions. In our case, we will use a fixed parametric form for $\beta(r)$, namely the Tiret model  $\text{''T''}$ of Ref. \cite{Tiret2007},
\begin{equation}\label{eq:betat}
\beta(r)_T=\beta_{\infty}\cdot \frac{1}{1+r_{c}/r} \, ,
\end{equation}
where the normalization of $\beta_\infty$ identifies the value of the anisotropy at large radii, while  $r_c$ is a characteristic scale radius.  The Tiret profile is a  generalized version of the Mamon and Lokas profile \cite{MamLok05}, which has been shown to provide a good fit to the average cluster-size halos anisotropy profile over a set of cosmological simulations (see e.g Ref. \cite{mamon10}). Here, we will assume that $r_c$ coincides with the scale radius of the mass profile $r_s$, and we will also fix $\beta_\infty=0.5$ to generate our sample.
With the above, the solution for the radial dispersion  $\sigma^2_r$  is given by \cite{MamLok05}
\begin{equation} 
\sigma^2_r(r)=\frac{1}{\nu(r)}\int_{0}^{\infty}\exp\left[2\int_{r}^{s}\beta(t)\frac{dt}{t}\right]\nu(s)\frac{d(\Psi(s))}{ds}ds \, , \label{eq:sigmar}
\end{equation}
which is connected to the angular one through 
\begin{equation}
\sigma^2_{\theta}(r)= [1-\beta(r)]\sigma^2_r(r) \, .
\end{equation}
Since we are assuming spherical symmetry we impose $\sigma^2_{\phi}\equiv\sigma^2_{\theta}$.

\subsection{Reconstructing the clusters dynamical mass: the \emph{MAMPOSSt} method}  \label{sec:MAM}
Armed with the previously constructed synthetic clusters,  each with a given $(r_{s}, r_{200})$ and a velocity distribution for its tracers, our next step is the reconstruction of what would be the cluster's mass profile from the kinematical data. 

To this purpose, we employ the {\it MAMPOSSt (Modeling Anisotropy and Mass Profiles of Observed Spherical Systems)} procedure developed by Ref. \cite{Mamon01}, which provides a numerical framework for deriving  the inferred cluster mass profiles from the analysis of the dynamics of the member galaxies. The code performs a maximum likelihood-fit to the distribution of the positions and velocity field of the galaxies in projected phase space $(R,v_z)$, where $R$ is the projected radius of each tracer from the cluster center and $v_z$ the velocity along the line of sight (los), in the cluster rest-frame. As an example, in the left panel of Figure \ref{fig:pps} we show the projected phase space of the $12^{th}$ synthetic cluster in the sample, generated with $1024$ particles within $r_{200}$. The radius $R=r_{200}$ is identified by the red vertical line, while the blue points indicate all the cluster members lying in projection within $r_{200}$.

Given a parametric form of the mass profile $M(r)$ and of the velocity anisotropy profile $\beta(r)=1-{\sigma^2_{\theta}}/{\sigma^2_r}$, the code solves the spherical Jeans equation \eqref{eq:jeans} to derive the probability density of finding an object at position $(R,v_z)$ in the projected phase space as
\begin{equation} \label{q}
q(R,v_z)=\frac{2\pi Rg(R,v_z)}{N_p(R_{max})-N_p(R_{min})}\, .
\end{equation}
Here $N_p(R)$ stands for the predicted cumulative number of objects at projected radius $R$ ,and $g(R,v_z)$ is the surface probability density of observed objects, which in the case of a 3D-Gaussian velocity distribution takes the form
\begin{equation}\label{eq:grv}
g(R,v_z)= \sqrt{\frac{2}{\pi}}\int_R^\infty\frac{r\nu(r)}{\sqrt{r^2-R^2}}\frac{dr}{\sigma_r(r)\sqrt{1-\beta(r)R^2/r^2}}\exp\left[-\frac{v^2_z}{2(1-\beta(r)R^2/r^2)\sigma^2_r(r)}\right] \, .
\end{equation}
Given eq. \eqref{q}, the likelihood is simply given by the sum over all the tracers in the projected phase space,
\begin{equation}\label{likelihood-Dyn}
-\log\mathcal{L}=-\sum_{i=1}^n\log q(R_i,v_{z,i}|\vec{\theta}) \, ,
\end{equation}
where as usual, $\vec{\theta}$ stands for the model-parameters vector. The version of \emph{MAMPOSSt} code used within this paper additionally requires a parametric model of the projected number density profile of the tracers
with a characteristic scale radius $r_{\nu}$. Since we are working with collisionless particles, by construction the number density profile $\nu(r)$ scales exactly as the NFW mass profile $\rho(r)$ (i.e. $r_{\nu}\equiv r_s$), however, the projected number density profile, needed to compute the probability of eq. \eqref{q}, is obtained in \emph{MAMPOSSt} by integrating the  the 3-dimensional profile along the line of sight, assuming that it extends to infinity. This leads to a value of $r_{\nu}$ which can be slightly different from $r_s$ and for this reason we fit the projected number density profile from the phase space of each synthetic cluster and use the best fit values of $r_{\nu}$ as the input for the \emph{MAMPOSSt} analysis (see the fourth column of Table \ref{tab:perfc}). This is also more reliable from an observational point of view:  in a real cluster it is not guaranteed that the distribution of the tracers (i.e. the member galaxies) scales in the same as the total density profile (see e.g. Refs. \cite{budzynski12,Biviano06}). 

\section{Results}\label{sec:results}
\begin{figure}
 \includegraphics[width=1.0\textwidth]{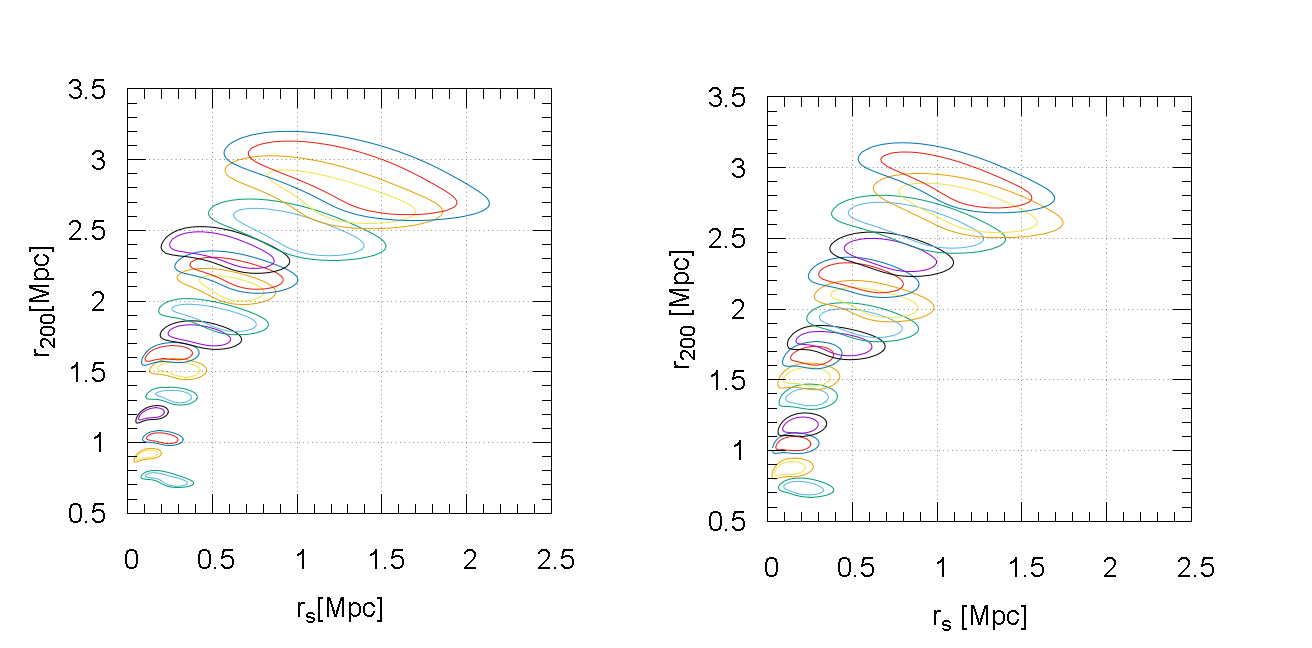}
\caption{\label{fig:dyn_ellips}$1\sigma$ and $2\sigma$ contours from the \emph{MAMPOSSt} analysis of the sample of synthetic clusters. The {\it left} plot shows the case where a fixed number of tracers is consider within $r_{200}$, while the {\it right} plot refers to the case where the number of tracers is proportional to the cluster mass (see Section \ref{sec:jeans}).}

\end{figure}

We now proceed with the results of our analysis aimed at constraining the gravitational slip $\eta$ from the sample of simulated synthetic clusters described in Section \ref{sec:Dynamics}. In particular, we will apply the \emph{MAMPOSSt} procedure to each halo in order to simultaneously constrain the mass profile parameters $(r_s,\,r_{200})$ and the velocity anisotropy parameter $\beta_{\infty}$, considering all the particles with a projected radius $R\in[0.05\,\mpc,\,r_{200}]$ in the fit. This choice has been made to be consistent with real observations, although the range on which we perform the fit could be arbitrary in this ideal case. Indeed, for an observed cluster we cannot assume the validity of the Jeans equation beyond $r_{200}$, while at very small radii the dynamics is dominated by the presence of the brightest central galaxy (BCG), which is the dominant
galaxy that sits at the bottom of the potential well of relaxed galaxy clusters (see e.g. Ref. \cite{Biviano01}). 
The final likelihood distribution of dynamics  $\mathcal{L}^{\text{dyn}} = \mathcal{L}^{\text{dyn}}(r_{200}^{\text{dyn}},r_{s}^{\text{dyn}})$ is then obtained by marginalizing over $\beta_\infty$. In Figure \ref{fig:dyn_ellips} we show the resulting $1\sigma$ and $2\sigma$ iso-probability contours for all the synthetic clusters obtained for what will be our {\it reference case} with $N_g(r_{200})=1000$ (left plot), and the case with a variable number of tracers within $r_{200}$ (right plot).
It is worth to point out that the best fit values $\bar{r}_s,\bar{r}_{200}$ given by the \emph{MAMPOSSt} analysis are not expected to coincide with the true values listed in Table \ref{tab:perfc}, even if these are always included within the  $1\sigma$ region. This fact is not surprising, since each synthetic projected phase space represents only one possible statistical realization of a halo with given mass, concentration and anisotropy (we stress again that velocities of the tracers are randomly generated over a Gaussian distribution with given  $\vec{\sigma}^2_{\mathbf{r}} (r)$).  
For the purposes of our analysis, we assume as fiducial model the GR expectation $\eta=1$, which is obtained by imposing that the lensing probability distribution $\mathcal{L}^{\text{len}} = P_L(r_{200}^{\text{len}},r_{s}^{\text{len}})$ peaks at the best fit values $\bar{r}_s,\bar{r}_{200}$ given by the dynamics,  through an appropriate rescaling which preserves  the relative errors
$(\sigma(r_s)/r_s)^{\text{lens}}$ and $(\sigma(r_{200})/r_{200})^{\text{lens}}$. 

%

\subsection{Case A: Scale--independent $\eta$}
The first part of the analysis considers $\eta$ to be constant at all scales. This implies that in eq. \eqref{eq:eta} the scale radius of the NFW profile should be the same as inferred by a lensing or dynamics probe, i.e. $r_s^{\text{lens}}=r_s^{\text{dyn}}\equiv r_s$. Thus, the expression of $\eta$ as a function of the  mass profiles $M^{\text{lens}},\,M^{\text{dyn}}$ reduces to a constant ratio where the only relevant parameters are the virial radii $r_{200}^{\text{len}},\, r_{200}^{\text{dyn}}$,
\be \label{eq:eta_si}
\eta=2\times\frac{\ln(1+c^{\text{dyn}}) - c^{\text{dyn}}/(1+c^{\text{dyn}}) }{\ln(1+c^{\text{lens}}) - c^{\text{lens}}/(1+c^{\text{lens}}) }\left(\frac{r_{200}^{\text{lens}}}{r_{200}^{\text{dyn}} }\right)^3-1 \, ,
\ee
where $c^{\text{lens/dyn}}=r_{200}^{\text{lens/dyn}}/r_s$ are the lensing and dynamics concentrations respectively.
Following the procedure of Ref. \cite{Pizzuti16}, each of the $15$ pairs of joint dynamics-lensing likelihoods are Monte-Carlo sampled for $10000$ sets of values $(r_s,\, r_{200}^{\text{dyn}}\,,r_{200}^{\text{lens}})$ to obtain the corresponding distribution for $\eta$ according to equation \eqref{eq:eta_si}. The analysis is iterated for the case of synthetic clusters produced with a fixed and a variable number of tracers respectively ($15$ clusters for each case), as explained in Section \ref{sec:jeans}. 

\begin{figure}
 \includegraphics[width=1.0\textwidth]{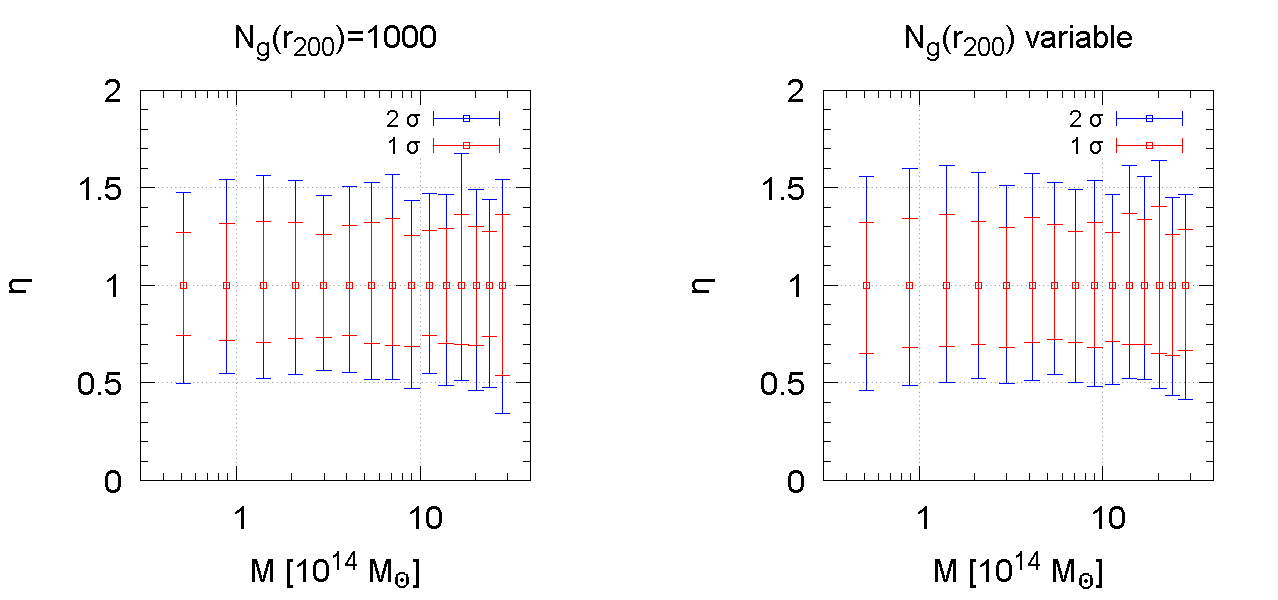}
\caption{$1\sigma$ and $2\sigma$ constraints on the gravitational slip parameter $\eta$ from the combination of galaxy cluster dynamics (simulated) and weak lensing data (MACS1206) for the sample of $15$ clusters considered in this work. The {\it left} plot corresponds to cluster simulations with a fixed number of tracers ($N_g=1000$), while the {\it right} plot to the case where the  number of tracers is proportional to the cluster mass (see Section \ref{sec:jeans}). It can be seen that, the constraints are practically insensitive to the values of mass and concentration of the cluster, as well as to the variation of the number of tracers according to cluster total mass.}
\label{plot:1case-single}
\end{figure}

\begin{figure}
\includegraphics[width=1.0\textwidth]{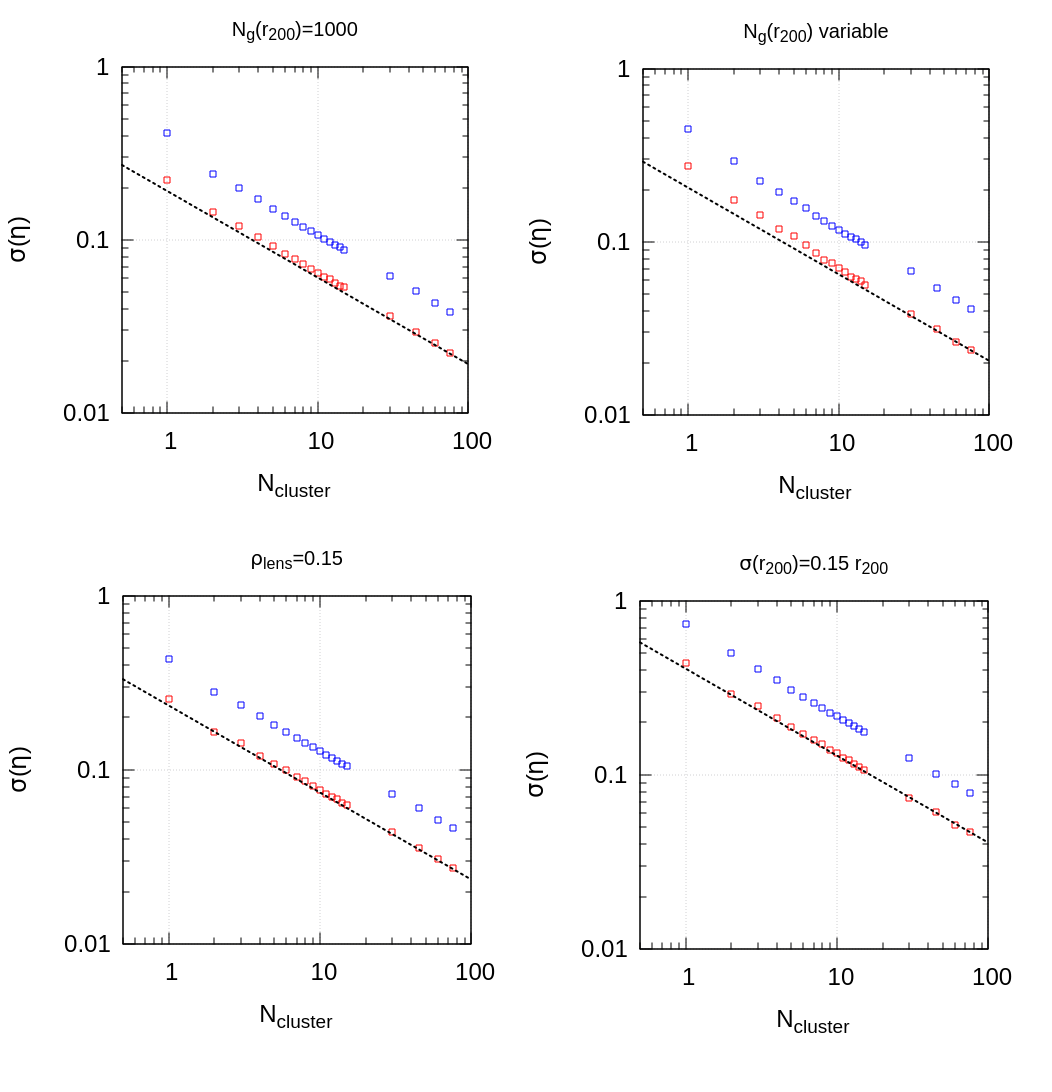} 
\caption{\label{plot:1case-all} $1\sigma$ (red, lower squares) and $2\sigma$ (blue, upper squares) constraints on the gravitational slip parameter $\eta$ from the combination of galaxy cluster dynamics (simulated) and weak lensing data (MACS1206) as described in the text. The last four data points are extrapolations of what would be the expected predictions based on the analysis of the $15$ clusters considered here. The {\it upper left} plot correspond to cluster simulations with a fixed number of tracers ($N_g=1000$), while the {\it upper right} one to the case where the tracers number is proportional to the cluster mass (see Section \ref{sec:jeans}). The {\it bottom left} plot shows the constraints obtained when changing the correlation of the lensing distribution from $\rho=0.67$ to $\rho=0.15$; finally, the {\it bottom right} plot is for the case when the standard deviation of $r_{200}$ is doubled. In all cases the fiducial model is GR, $\eta = 1$. The black-dotted curve denotes the theoretically expected scaling of the $1\sigma$ errors, $\sim \left(N_{\text{cluster}}\right)^{-1/2}$, which can be seen to be asymptotic with increasing $N_{\text{cluster}}$. Comparison of the two upper plots shows that the predicted errors appear to be essentially insensitive to the choice of the number of tracers in the cluster.}
\end{figure}

The results are shown in Figures \ref{plot:1case-single} and \ref{plot:1case-all} . In particular, Figure \ref{plot:1case-single} shows the $1\sigma$ and $2\sigma$ constraints on $\eta$ from each individual cluster, while Figure \ref{plot:1case-all} shows the combined constraints from many clusters. For $N>15$ we extrapolate the predictions on $\eta$ by naively multiplying several times the 15 cluster-combined distribution.
An important feature shown in Figure \ref{plot:1case-single} is that the constraints on $\eta$ are practically insensitive to the choice of the member galaxies used within the \emph{MAMPOSSt} fit, masses and concentration of the cluster, in agreement with the discussion of Ref. \cite{Mamon01}. From an observational viewpoint, this is particularly encouraging, since it broadens the range of clusters that may be used from current or future surveys. 
Furthermore, from Figure \ref{plot:1case-all} it can be seen that the error on $\eta$ obtained by combining together the distributions of many clusters ($N_{\rm clusters}$) scales approximately as the expected theoretical behavior $\sim (N_{clusters})^{-1/2}$ (black dashed line in all panels). We predict that, for our reference analysis (top left panel) the {\it combined analysis of 15 clusters} lead to a constraint 
\be
\eta=1.00\pm 0.05 \; \;  \text{at} \; \; 1 \sigma \, , 
\ee
which corresponds to an average statistical uncertainty of $5.5\%$. Furthermore, it turns out that with {\it 75 clusters} we reach the accuracy of $\sim 2\%$ and $\sim 4\%$ at $1\sigma$ and $2 \sigma$ respectively. When varying the number of particles in the dynamics fit (top right panel) the scaling relation mildly deviates from $(N_{\textrm{clusters}})^{-1/2}$, as a consequence of the slightly larger uncertainties on $\eta$ from the analysis of the smallest halos. The effect is however averaged out when increasing $N_{\rm clusters}$ and the combined constraints are identical to the reference case.

In order to investigate the effect on our predictions induced by variations in the lensing information (still relying on the simplified assumption of a Gaussian distribution) we repeat the analysis doubling the uncertainties in $r_{200}$ and changing the value of the correlation $\rho$ in the lensing information. We have considered three alternatives for $\rho$: a totally uncorrelated distribution ($\rho=0$), a small correlation $\rho=0.15$ and  $\rho=0.99$.
The results are shown in the lower panels of Figure \ref{plot:1case-all} (for the correlation we plot only the cae $\rho=0.15$) . For completeness, it is worth noticing that we have also considered a change in the relative uncertainties of the scale radius $\sigma(r_{s})/r_s$ from $0.3 \to 0.45$, but the variation has only marginal effects on the final distributions. 
The marginalized constraints on $r_s$ obtained by the \emph{MAMPOSSt} fit on our synthetic sample are on average  of the same order of the one used in the lensing distribution for the reference analysis $\sigma(r_{s})/r_s\sim 0.3$. Therefore, modifying the lensing information on $r_s$ in practice does not affect the final predictions for $\eta$. We nevertheless emphasize that mild effects can be enhanced or suppressed depending on the adopted value of the correlation $\rho$. All the constraints obtained in this case are listed in the second column of Table \ref{tab:res}.

\subsection{Case B: Scale--dependent $\eta$}
We now consider $\eta\equiv\eta(r)$ to be scale--dependent, i.e. the scale radius of the lensing mass profile ($r_s^{L}$), and that of the dynamics mass profile ($r_s^{D}$) are independent parameters for each cluster. As shown in Ref. \cite{Pizzuti16}, the constraints on $\eta$ depend on the distance from the cluster center, with the scaling of the uncertainties with $r$ determined by the actual values of $r_s^L,\, r_s^D$. 
In order to combine the distributions of $\eta$ obtained for each single cluster we considered 100 equally-spaced radial bins in the range $[0.1,1.5]\,\mpc$. The choice of the lower bound is dictated by consistency with real observations -- indeed, as already explained earlier, the assumption of the Jeans analysis cannot be met at very small radii where the dynamics is expected to be dominated by dissipative effects. As for the upper bounds, for an observed cluster the analysis cannot generally be performed beyond the virial radius $r>r_{200}$, where lack of dynamical equilibrium as well as contamination from the large scale structure introduces systematic effects in both lensing and dynamics mass profile reconstructions. 
We considered the value $r_{max}=1.5\,\mpc$ as a fair compromise that accounts for the different scales of the mass profiles.  
In each radial bin we Monte-Carlo sampled the joint probability distribution $(r_s^L,r_{200}^L,rs_D,r_{200}^D)$ for 10000 trials, and combined the resulting likelihoods for $\eta(r)$ to get the final probability distribution $P(\eta,r)$. As before, for $N_{cluster}>15$ we extrapolated the behaviour of the constraints by simply multiplying different times $P(\eta,r)$ by itself. 

\begin{figure}
 \includegraphics[width=1.0\textwidth]{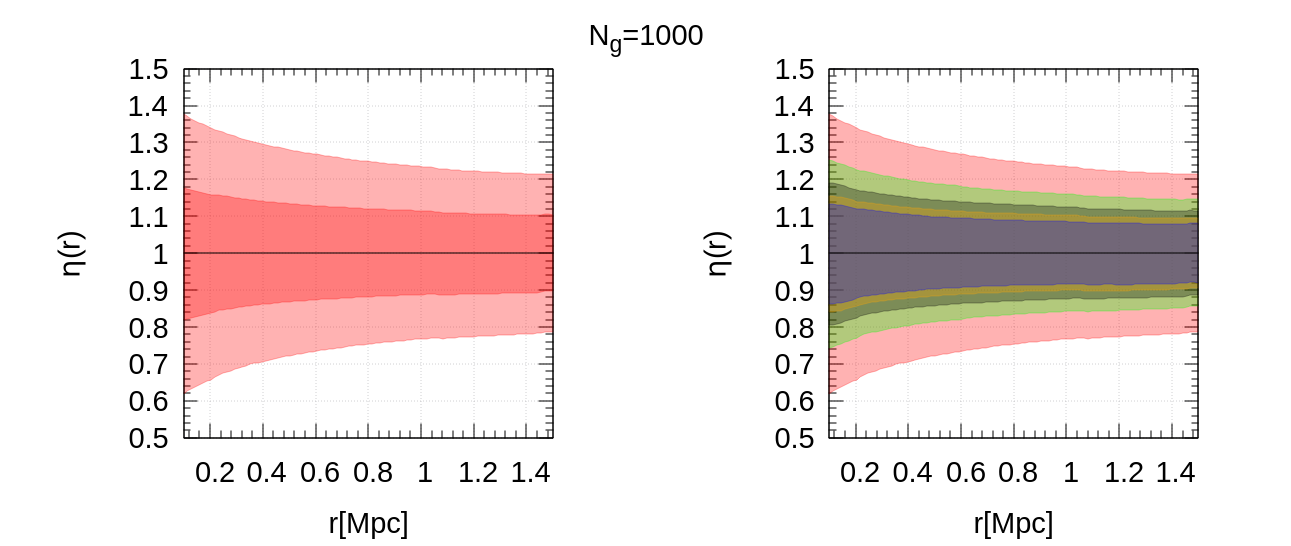}
\caption{\label{plot:2case-Ng1000}Constraints on a scale-dependent $\eta$ as a function of the radius from the center of the cluster for the case of a constant number of member galaxies ($N_g = 1000$) throughout our synthetic cluster sample. {\it Left:} $1\sigma$ and $2\sigma$ bands accounting for the sample of $15$ synthetic clusters, as explained in the text. {\it Right:} $2\sigma$ bands derived from different numbers of clusters as follows (from top to bottom): Red $ = 15$, Green $= 30$, Dark Green $= 45$, Yellow $=60$ and Violet $= 75$ clusters respectively. The cases with a number of clusters $> 15$ are derived by naively multiplying the combined likelihood of $15$ clusters by $n = \{2, 3, \ldots \}$.}

\end{figure}

\begin{figure}
 \includegraphics[width=1.0\textwidth]{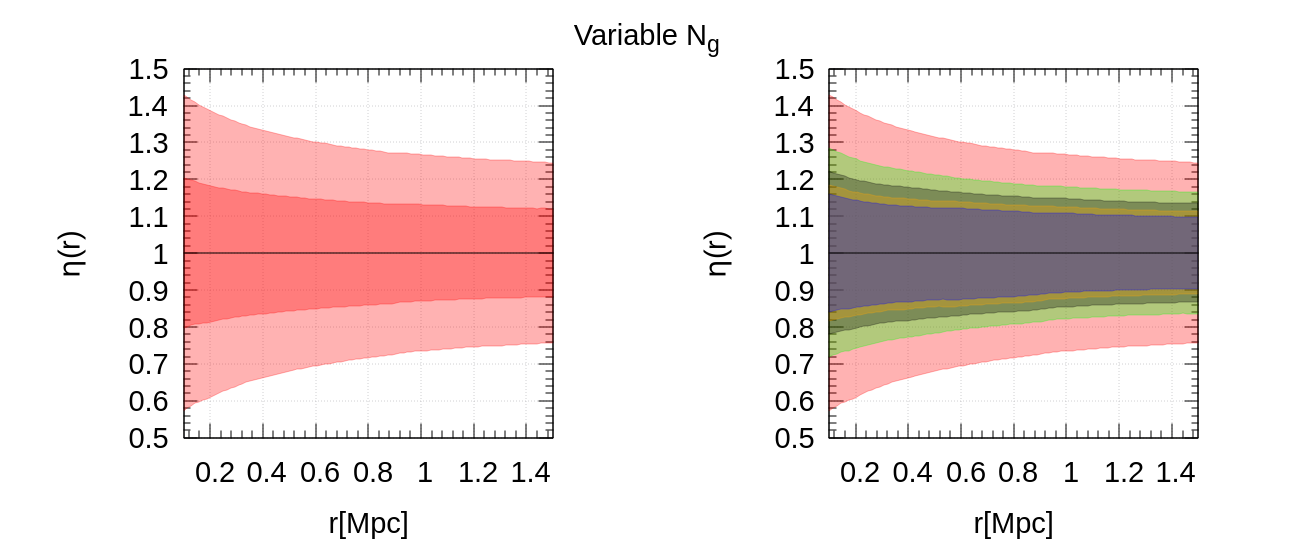}
\caption{\label{plot:2case-VarNg}Similar to Figure \ref{plot:2case-Ng1000}, but now the synthetic clusters used in the analysis are constructed with a variable number of member galaxies proportional to the cluster mass (see Section \ref{sec:jeans}). In particular, for the case of $75$ clusters it turns out that $\eta( r = 1.5 \, \text{Mpc}) =1 \pm 0.04$ at $1 \sigma$ and $\eta( r = 1.5 \, \text{Mpc}) =1 \pm 0.1$ at $2 \sigma$ respectively.}

\end{figure}

In Figures \ref{plot:2case-Ng1000}, \ref{plot:2case-VarNg} we show the results for the reference analysis of $N_g = 1000$, and for the case of variable number of tracers in the \emph{MAMPOSSt} fit respectively. In each plot the left panel refers to one (inner red shaded area) and two (outer red shaded region) $\sigma$ bands for the 15 cluster-combined distributions while the right panel shows the scaling of the $2\sigma$-errors when increasing $N_{cluster}$ from 15 (red, outer) to 75 (gray, inner). The constraints for all the sub-cases analysed in the scale-dependent configuration, evaluated at the radius $r=1.5\,\mpc$, are summarized in  the third column of Table \ref{tab:res}. As now $r_s^D\neq r_s^L$,  we consider also the variation of the uncertainties in the lensing scale radius $r_s^L$ from 30\% to 45\% which produces an observable effect on the bounds on $\eta$.

\begin{table}
\centering
\begin{tabular}{c|c|c|c}
\hline
\bf{case} & ${N_{cluster}}$ &  $\eta = \text{constant}$ & $\eta = \eta(r = 1.5 \text{Mpc})$   \\ \hline
\hline
&&&\\
reference & 15  &	$1.00\pm 0.05\,\,(1\sigma) \pm 0.09 \,\,(2\sigma)$  & $1.00^{+0.10}_{-0.09}\,\,(1\sigma) \pm 0.21\,\,(2\sigma)$ \\	
 & 75  & $1.00 \pm 0.02\,\,(1\sigma) \pm 0.04 \,\,(2\sigma) $ & $1.00^{+0.04}_{-0.03}\,\,(1\sigma)\pm0.08\,\,(2\sigma)$ \\ \hline \hline
&&&\\
Variable $N_g$ & 15 &	$1.00 \pm 0.05\,\,(1\sigma) \pm 0.09 \,\,(2\sigma) $  & $1.00 \pm 0.12\,\,(1\sigma)  {}^{+0.25}_{-0.24}\,\,(2\sigma)$ \\	
& 75   & $1.00 \pm 0.02\,\,(1\sigma) \pm 0.04 \,\,(2\sigma) $ & $1.00 \pm 0.04\,\,(1\sigma) {}^{+0.10}_{-0.09}\,\,(2\sigma)$\\ \hline \hline
&&&\\
$\rho=0.15$ & 15 &     $1.00 \pm 0.06\,\,(1\sigma), \pm 0.11\,\,(2\sigma)$  & $1.00 \pm 0.13\,\,(1\sigma), \pm 0.27\,\,(2\sigma)$ \\	
& 75  & $1.00 \pm 0.03\,\,(1\sigma), \pm 0.05\,\,(2\sigma)$ & $1.00 \pm 0.05\,\,(1\sigma) \pm 0.09\,\,(2\sigma)$\\ \hline \hline
&&&\\

$\rho=0$ & 15 &     $1.00 ^{+0.06}_{-0.05}\,\,(1\sigma), \pm 0.10\,\,(2\sigma)$  & $1.00 \pm 0.12\,\,(1\sigma), \pm 0.26\,\,(2\sigma)$ \\	
& 75  & $1.00 ^{+0.03}_{-0.02}\,\,(1\sigma), \pm 0.04\,\,(2\sigma)$ & $1.00 \pm 0.05\,\,(1\sigma) \pm 0.08\,\,(2\sigma)$\\ \hline \hline
&&&\\

$\rho=0.99$ & 15 &     $1.00 \pm 0.03\,\,(1\sigma), \pm 0.11\,\,(2\sigma)$  & $1.00 \pm 0.08\,\,(1\sigma), \pm 0.19\,\,(2\sigma)$ \\	
& 75  & $1.00 \pm 0.01\,\,(1\sigma), \pm 0.03\,\,(2\sigma)$ & $1.00 \pm 0.03\,\,(1\sigma) \pm 0.06\,\,(2\sigma)$\\ \hline \hline
&&&\\

$\sigma( r_{200})=0.15\times r_{200}$ & 15 &	$1.00^{+0.11}_{-0.10}\,\,(1\sigma) {}^{+0.18}_{-0.17}\,\,(2\sigma)$  & $1.00 \pm 0.15\,\,(1\sigma) \pm 0.30\,\,(2\sigma)$ \\	
& 75  & $1.00 \pm 0.05\,\,(1\sigma) \pm 0.08\,\,(2\sigma)$ & $1.00 \pm 0.05\,\,(1\sigma) \pm 0.11\,\,(2\sigma)$\\ \hline \hline
&&&\\
$\sigma( r_{s})=0.45\times r_{s}$ & 15 & $1.00 \pm 0.05\,\,(1\sigma) \pm 0.09 \,\,(2\sigma) $  & $1.00 \pm 0.15\,\,(1\sigma) \pm 0.31\,\,(2\sigma)$ \\	
& 75   &$1.00 \pm 0.02\,\,(1\sigma) \pm 0.04 \,\,(2\sigma) $ & $1.00 \pm 0.05\,\,(1\sigma) \pm 0.12\,\,(2\sigma)$\\ \hline \hline
\end{tabular}
\caption{\label{tab:res} Constraints from the combined analysis of multiple number of clusters for scale-independent (third column) and  scale-dependent (fourth column) gravitational slip $\eta$. In the latter case, the error is evaluated at the reference scale $r = 1.5\,\text{Mpc}$. In each block each of the two rows refer to a different sub-case: the first one shows the constraints from the 15 clusters-likelihood, while the second refers to the extrapolated constraints for $N_{cluster}=75$. When varying the lensing parameters (correlation $\rho$ and relative errors respectively), the number of tracers in the dynamics fit is fixed to $N_g=1000$ for all clusters.}
\end{table}

\subsection{The effect of the density of member galaxies} \label{sec:clusterdens}
Realistic surveys cover a wide range of clusters, each with a different number of tracers depending on the details of its evolution and environment. In this regard, an important question regards the dependence of constraints on the density of tracers within the cluster. Here, we will provide an answer considering the case of a scale-independent $\eta$, and we will show that the effect is practically negligible above a given density of tracers. For this purpose, we generated 6 realisations of a new cluster, characterised by fixed $r_{200}=1.96\,\mpc$, $r_s=0.27\, \mpc$, and a Tiret velocity anisotropy profile with $\beta_\infty=0.5$, increasing the number of particles enclosed within $r_{200}$ in each realisation, considering the cases $N_g(r_{200})=50, 100, 200, 500, 1000, 2000$. Notice that we chose to keep the lensing probability distribution identical to the reference case. 

\begin{figure} 
 \includegraphics[width=1.0\textwidth]{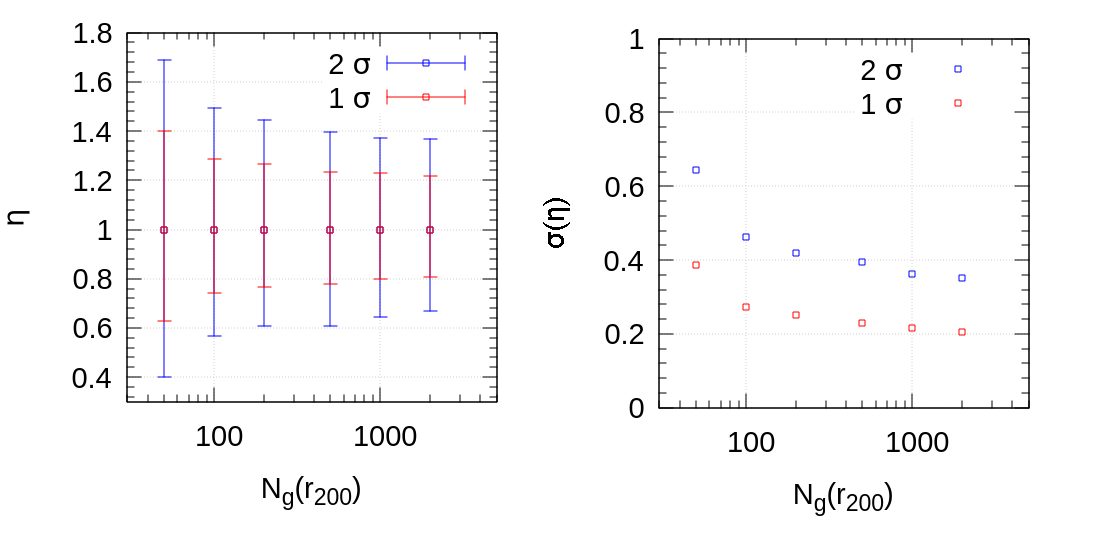}
\caption{\label{fig:varydens} Constraints on a scale-independent $\eta$ as a function of the number of tracers considered within $r_{200}$ of a single cluster, as explained in Section \ref{sec:clusterdens}. {\it Left panel:} $1 \sigma$ and $2 \sigma$ errorbars. {\it Right panel:} Scaling of the $1 \sigma$ and $2 \sigma$ symmetrised errors with the tracers number.}
\end{figure}

In Figure  \ref{fig:varydens} we show the $1 \sigma$ and $2 \sigma$ error bars  (left panel), along with the symmetrised $1 \sigma$ and $2 \sigma$ errors (right panel) of  a scale--independent $\eta$ as a function of $N_g(r_{200})$. It appears that above $\sim100$ tracers within $r_{200}$, the goodness of the resulting bounds on $\eta$ increases only marginally, passing from $27\%$  uncertainties at $1 \sigma$ for $N(r_{200})= 100$, to  $21\%$ for $N(r_{200})= 2000$; the effect is almost irrelevant for $N(r_{200})>500$. 

\begin{figure}
\centering
 \includegraphics[width=0.55\textwidth]{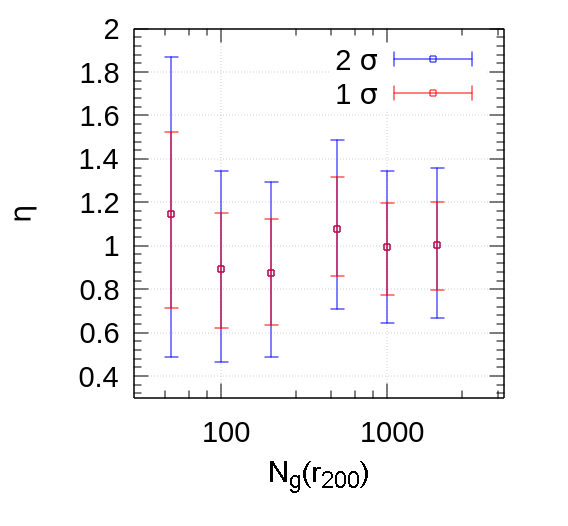}
\caption{\label{fig:increase_nobf} Similar to Figure \ref{fig:varydens}, but with the simulated lensing distribution now centered on the true values of the cluster mass profile parameters. Notice that, the fiducial is not centered around $\eta = 1$ anymore.  }

\end{figure}

Extrapolating the result of $\eta(N_g=100)$ with the naive theoretically expected scaling $\sim N_{cluster}^{-1/2}$, we obtain 
\begin{align}
& \eta(N_g=100)=1.00\pm 0.07 (1 \sigma) {}^{+0.13}_{-0.11} (2 \sigma)\,\,\, \text{15 clusters}, \\ 
& \eta(N_g=100)=1.00\pm 0.03 (1 \sigma){}^{+0.06}_{-0.05} (2 \sigma)\,\,\, \text{75 clusters}.
\end{align}
This shows that very stringent constraints on the gravitational slip can be obtained with a reasonably small number of clusters for which a hundred measured spectroscopic redshifts are available, together with reliable lensing mass profile reconstructions. It is worth pointing out that the analysis carried out in this paper relies on the assumption that the best fit values of $r_s,\,r_{200}$ from the simulated lensing distribution always coincides with the \emph{MAMPOSSt} best fit in order to guarantee that the resulting distribution of $\eta$ is centered over the GR expectation $\eta=1$. However, one should take into account that, even in this ideal situation where all the systematics are neglected, the smaller is the number of tracers in the \emph{MAMPOSSt} analysis, the larger is the difference between the true value of the mass profile parameters and the one fitted by \emph{MAMPOSSt} (the so called bias parameter, see e.g. Ref. \cite{Mamon01}). In Figure \ref{fig:increase_nobf} we plot the error bars on $\eta$ obtained by centering the lensing distribution on the true values $r_{200}=1.96\,\mpc,r_{s}=0.27\,\mpc$ instead of on the \emph{MAMPOSSt} best fit values. As expected, the bias in $\eta$ tends to be larger for smaller number of tracers, even if GR is always included within $1\sigma$.

\section{Summary and discussion}
The gravitational slip parameter $\eta \equiv \Phi/\Psi$ is of paramount importance for tests of gravity at cosmological and astrophysical scales. While a possible future detection of $\eta \neq 1$ would be the smoking gun for a departure from the standard gravity paradigm of GR at late-times in the cosmological evolution, a non-detection would disfavor large families of modified gravity scenarios such as the popular $f(R)$ scalar-tensor theories. 
Here, we elaborated on the potential of galaxy cluster observations to constrain $\eta$, through a joint analysis of dynamics and lensing information. As we discussed in detail in Section \ref{sec:jeans}, our ground for the analysis of dynamics of cluster galaxies was based on a sample of synthetic (simulated) clusters, whose dynamical mass profile were subsequently derived by means of the \emph{MAMPOSSt} numerical framework, in turn to be combined with the lensing one using the available information from the analysis of the cluster MACS 1206. 

It is important to remind that our reconstruction of $\eta$ did not invoke the assumption of a particular gravity models or parametrisation. In our view, $\eta \neq 1$ is sourced through a difference in the respective values of the NFW parameters $(r_{200}, r_s)$ when inferred from dynamics and lensing respectively. This phenomenology inclined view is  general and allows to translate the derived constraints on $\eta$ to constraints on the theory space of specific models. Nevertheless, we inevitably had to rely on assumptions such as virialization, spherical symmetry (see below) and a NFW parametrisation for the matter density. In addition, the analysis was performed at a fixed redshift $z = 0$, which prevented us to capture any possible time-dependence of $\eta$. As concerns the lensing distribution, it is worth to notice that we have performed our analysis under the simplified assumption of Gaussianity for the lensing probability distribution $P_L(r_s,r_{200})$ (even if well-motivated by the reference observational data of MACS 1206). The effect of a variation of the lensing information such as the relative errors and correlation was investigated to find that it has a mild, yet sizable effect on the derived constraints as can be seen from Table \ref{tab:res}.

We can summarise our {\it main results} as follows: 
\\ \\ 
1. For a scale-independent $\eta$, and assuming a fiducial with $\eta = 1$, we can reach an accuracy of $5 \%$ ($9 \%$) at $1\sigma$ ($2\sigma$) with $15$ clusters, which is brought down to $2 \%$ and $4\%$ respectively when extrapolated to $75$ clusters. The explicit results are also summarized in Table \ref{tab:res}, and Figures \ref{plot:2case-Ng1000} and \ref{plot:2case-VarNg}.
\\ \\
2. In a similar manner, for the more realistic case of a scale-dependent $\eta = \eta(r)$, when evaluated at the reference scale $r=1.5 \, \text{Mpc}$ and accounting for $15$ clusters we find an accuracy of $~ 10 \%$ at $1\sigma$ and $21\%$ at $2\sigma$, which decreases to $\sim 4\%$ and $8\%$ when extrapolating to $75$ clusters. Of course, it should come as no surprise that the constraints are less optimistic compared to the simpler scale-independent case.
\\ \\ 
3. Optimisation of future data analyses in this context requires an understanding of the dependence of the constraints on factors such as the clusters masses and the number of galaxy members in the clusters. Our investigation revealed that the cluster mass (parametrised through $r_{200}$ in the NFW profile) has a rather mild effect on $\eta$. This is best seen from the constraints based on the individual clusters as a function of the cluster mass, as it is shown in Figure \ref{plot:1case-single}. According to the procedure explained in Section \ref{sec:jeans} and shown in Table \ref{tab:res}, varying the number of tracers in the clusters according to the halo total mass has either none or a very mild quantitative effect on the constraints. To better understand this behavior, we have further investigated the effect of increasing the density of tracers used in the dynamics fit for one single cluster. The results of this exercise show that after $N_g(r_{200})\sim 10^2$ the effect on the constraints for a scale--independent $\eta$ is moderate, becoming irrelevant for $N_g(r_{200})\gtrsim 500$. (See Figures \ref{fig:varydens} and \ref{fig:increase_nobf}.)
\\

This paper has been devoted to the study of the constraining power of galaxy cluster mass profiles as a probe for modified gravity. It is worth stressing out again that the analysis carried out addressed the problem only from a statistical point of view, neglecting the effect of possible systematics. In principle, several issues can introduce biases in the determination of the gravitational slip --  for example, deviations from spherical symmetry and dynamical relaxation, which are the main assumptions of our method, can be sources of differences between the mass profile inferred from lensing and dynamics of the member galaxies respectively, giving rise to spurious detections of gravitational slip, $\eta\neq 1$. 
Moreover, as already mentioned in Section \ref{sec:jeans}, the velocity anisotropy profile $\beta(r)$ is generally an unknown in analyses of cluster dynamics. In the parametric approach implemented in our version of the \emph{MAMPOSSt} code, the choice of the anisotropy model introduces additional systematics that should be taken into account. A similar statement can be made concerning the parametrisation of the mass profile, as discussed in Ref. \cite{Pizzuti16}. An additional source of systematics not considered here is the presence of interlopers in the dynamical analysis, namely galaxies that are projected in the cluster area but are not dynamically bound to the cluster.

The large amount of data for several hundred clusters that will be provided by current and future imaging and spectroscopic surveys, can allow for determinations of the gravitational slip parameter down to the percent level. This naturally calls for an accurate calibration of the impact of systematics and understanding of the relevant statistics. Such constraints are complementary to other cosmological probes such as the Cosmic Microwave Background, gravitational waves and electromagnetic counterpart detections, cluster number counts and redshift space distortions, allowing to test gravity at cosmological scales down to an unprecedented level of accuracy.

\section*{Acknowledgments}
I.D.S is supported by ESIF and MEYS (Project CoGraDS --  CZ.02.1.01/0.0/0.0/15\_003/0000437).
S.C. acknowledges grants by CNES and CNRS. L.P. acknowledges S. Borgani and B. Sartoris for useful comments and discussions.
\bibliographystyle{JHEP}
\bibliography{master, AnisoRefs, WD-Bib}

\end{document}